\documentclass[12pt]{article}
  \usepackage{amsfonts}
  \usepackage{amsmath}
\usepackage{amssymb}
\usepackage{amscd}
\usepackage[dvips]{graphicx}
\begin{document}
~~
\bigskip
\bigskip
\begin{center}
{\Large {\bf{{{Generalized twist deformations of Poincare and
Galilei quantum groups}}}}}
\end{center}
\bigskip
\bigskip
\bigskip
\begin{center}
{{\large ${\rm {Marcin\;Daszkiewicz}}$ }}
\end{center}
\bigskip
\begin{center}
{ {{{Institute of Theoretical Physics\\ University of Wroc{\l}aw pl.
Maxa Borna 9, 50-206 Wroc{\l}aw, Poland\\ e-mail:
marcin@ift.uni.wroc.pl}}}}
\end{center}
\bigskip
\bigskip
\bigskip
\bigskip
\bigskip
\bigskip
\bigskip
\bigskip
\begin{abstract}
The three quantum groups dual to the generalized twist deformed
Poincare Hopf algebras are provided with use of FRT procedure. Their
 Galilean  counterparts are obtained by
nonrelativistic contraction scheme.
\end{abstract}
\bigskip
\bigskip
\bigskip
\bigskip
\bigskip
\bigskip
\bigskip
\bigskip
\bigskip
 \eject

\section{Introduction}

There are two major approaches to describe the particle kinematics
in  high energy (transplanckian) regime. The first treatment assumes
that relativistic symmetry becomes broken close to the Planck scale,
so that effectively we have to do with some preferred (cosmic) frame
(see  \cite{violating} and references therein). In the second
approach one assumes that Poincare symmetry is still present, but
together with the corresponding space-time it becomes deformed. The
second treatment follows from many phenomenological (see \cite{1a},
\cite{1b}) as well as from formal arguments, based mainly on Quantum
Gravity \cite{2}-\cite{2a} and String Theory models \cite{recent},
\cite{string1}.

It is well known that  a proper modification of the Poincare and
Galilei Hopf algebras can be realized in the framework of Quantum
Groups \cite{qg1}-\cite{qg3}. Hence, in accordance with general
Hopf-algebraic classification of all possible deformations of
relativistic and nonrelativistic symmetries \cite{zakrzewski},
\cite{kowclas},  one can distinguish three basic  kinds of quantum
spaces:\\
\\
{ 1)} Canonical ($\theta^{\mu\nu}$-deformed) space-time
\begin{equation}
[\;{\hat x}_{\mu},{\hat x}_{\nu}\;] = i\theta_{\mu\nu}\;\;\;;\;\;\;
\theta_{\mu\nu} = {\rm const}\;, \label{noncomm}
\end{equation}
considered in  \cite{3a}-\cite{3c}. The corresponding twist
deformation of  Poincar{e} Hopf algebra $\,\mathcal{U}_\theta({P})$
has been proposed in \cite{3a}, while its dual quantum group
$\mathcal{P}_{\theta}$  in \cite{3e} and \cite{3c}. There were also
provided   two $\theta^{\mu\nu}$-deformed Galilei Hopf algebras
\cite{dasz} as the contraction limits of twisted  Poincar{e} group
$\,\mathcal{U}_\theta({P})$. Their dual Hopf
structures have been represented in \cite{dualdasz}.\\
\\
{ 2)} Lie-algebraic modification of classical space
\begin{equation}
[\;{\hat x}_{\mu},{\hat x}_{\nu}\;] = i\theta_{\mu\nu}^{\rho}{\hat
x}_{\rho}\;, \label{noncomm1}
\end{equation}
with  particularly chosen coefficients $\theta_{\mu\nu}^{\rho}$
being constants. There exist two explicit realizations of such a
noncommutativity - $\kappa$-Poincar{e} Hopf algebra
$\,\mathcal{U}_\kappa({P})$  \cite{4a}, \cite{4b} and  twisted
Poincar{e}  Hopf structure $\,\mathcal{U}_\zeta({P})$ \cite{lie2}
(see also \cite{lie1}). Their dual partners $\mathcal{P}_\kappa$ and
$\mathcal{P}_\zeta$ have been discovered in \cite{zakrzewskigrupa}
and \cite{lie2}, respectively. Besides, the so-called
$\kappa$-Galilei
 group has been provided by nonrelativistic
contraction of $\kappa$-Poincar{e}  Hopf algebra in \cite{kappaga},
and its dual quantum partner has been described in \cite{gg}. The
remaining Galilei  algebras and corresponding quantum groups were
discovered in \cite{dasz} and \cite{dualdasz} by various contraction
schemes   of
twisted Poincar{e}  Hopf structure
 $\,\mathcal{U}_\zeta({P})$ and its dual partner $\mathcal{P}_\zeta$.\\
\\
{ 3)} Quadratic deformation of Minkowski space
\begin{equation}
[\;{\hat x}_{\mu},{\hat x}_{\nu}\;] =
i\theta_{\mu\nu}^{\rho\tau}{\hat x}_{\rho}{\hat x}_{\tau}\;,
\label{noncomm2}
\end{equation}
with coefficients $\theta_{\mu\nu}^{\rho\tau}$ being constants. This
type of noncommutativity has been proposed in \cite{lie2}  as the
translation sector of   Hopf structure $\mathcal{P}_\xi$. The
explicit form of its
nonrelativistic counterpart has been provided in \cite{dualdasz}. \\

Recently, there was considered other type of quantum space -
so-called generalized quantum space-time
\begin{equation}
[\;{\hat x}_{\mu},{\hat x}_{\nu}\;] = i\theta_{\mu\nu} +
i\theta_{\mu\nu}^{\rho}{\hat x}_{\rho}\;, \label{general}
\end{equation}
which combines   canonical type with the Lie-algebraic kind  of
space-time noncommutativity. The Hopf-algebraic realization of
corresponding quantum symmetries has been proposed in
\cite{lulya}-\cite{genpogali} in the case of relativistic symmetry
and in \cite{genpogali} as well for its nonrelativistic counterpart.

In this article we provide three Poincare quantum groups ${\mathcal
P}_{\theta_{kl},\kappa}$, ${\mathcal P}_{\theta_{0i},\hat{\kappa}}$
and ${\mathcal P}_{\theta_{0i},\bar{\kappa}}$ dual to the twist
deformed (generalized) Hopf  universal enveloping algebras
$\,\mathcal{U}_{\theta_{kl},\kappa}(P)$,
$\,\mathcal{U}_{\theta_{0i},\hat{\kappa}}(P)$ and
$\,\mathcal{U}_{\theta_{0i},\bar{\kappa}}(P)$ proposed in
\cite{genpogali} (see Section 2). All of them are obtained by
so-called FRT procedure \cite{frt}. Besides, we find their three
Galileian counterparts with use of the well known nonrelativistic
contraction scheme \cite{cont1}-\cite{cont3}.

It should be noted that obtained in such a way Galilei Hopf
structures can be get by direct application of FRT procedure as
well. However, the contraction scheme used in this article  has one
advantage - it gives more precise  information about investigated
objects, i.e. about the twisted Poincare and Galilei quantum groups
as well as about the linking contraction between both Hopf
structures. The relations between different types of Hopf algebras
and corresponding dual quantum groups are illustrated on Figure 1.

The paper is organized as follows. In second section we recall basic
facts concerning the relativistic
$\,\mathcal{U}_{\theta_{kl},\kappa}(P)$,
$\,\mathcal{U}_{\theta_{0i},\hat{\kappa}}(P)$,
$\,\mathcal{U}_{\theta_{0i},\bar{\kappa}}(P)$ and nonrelativistic
$\,\mathcal{U}_{\xi_{kl},\lambda}(G)$,
$\,\mathcal{U}_{\xi_{0i},\hat{\lambda}}(G)$,
$\,\mathcal{U}_{\xi_{0i},\bar{\lambda}}(G)$ Hopf algebras
respectively. In section three we provide corresponding six (dual)
quantum groups - three at relativistic level obtained by use of FRT
procedure and three at nonrelativistic level by the use of
contraction scheme. The final remarks are discussed in the last
section.

\section{Generalized twist deformations of Poincare and
Ga-
lilei Hopf algebras}

\subsection{Relativistic case}

In this section, following the paper \cite{genpogali}, we recall
basic facts related with  the generalized twist-deformed Poincare
Hopf algebras $\,\mathcal{U}_{\theta_{kl},\kappa}(P)$,
$\,\mathcal{U}_{\theta_{0i},\hat{\kappa}}(P)$ and
$\,\mathcal{U}_{\theta_{0i},\bar{\kappa}}(P)$. All of them are
described by so-called Abelian $r$-matrices $r_{\cdot,\cdot}\in
{\mathcal U}_{\cdot,\cdot}(P) \otimes \, {\mathcal
U}_{\cdot,\cdot}(P)$, which satisfy the classical Yang-Baxter
equation (CYBE)
\begin{equation}
[[\;r_{\cdot,\cdot},r_{\cdot,\cdot}\;] ] = [\;r_{\cdot,\cdot
12},r_{\cdot,\cdot13} + r_{\cdot,\cdot 23}\;] + [\;r_{\cdot,\cdot
13}, r_{\cdot,\cdot 23}\;] = 0\;, \label{cybe}
\end{equation}
where   symbol $[[\;\cdot,\cdot\;]]$ denotes the Schouten bracket
and for $r_{\cdot,\cdot} = \sum_{i}a_i\otimes b_i$
$$r_{\cdot,\cdot 12} = \sum_{i}a_i\otimes b_i\otimes 1\;\;,\;\;r_{\cdot,\cdot 13} = \sum_{i}a_i\otimes 1\otimes b_i\;\;,\;\;
r_{\cdot,\cdot 23} = \sum_{i}1\otimes a_i\otimes b_i\;.$$ Becouse
the classical $r_{\cdot,\cdot}$-matrices are spaned by Abelian
algebra, the corresponding twist factors are given by (see
\cite{qg1}-\cite{qg3})
\begin{eqnarray}
{\cal F}_{\cdot,\cdot} = \exp \left(ir_{\cdot,\cdot}\right)\;.
\label{factors}
\end{eqnarray}
They satisfy the classical cocycle condition
\begin{equation}
{\mathcal F}_{{\cdot,\cdot }12} \cdot(\Delta_{0} \otimes 1) ~{\cal
F}_{\cdot,\cdot } = {\mathcal F}_{{\cdot,\cdot }23} \cdot(1\otimes
\Delta_{0}) ~{\mathcal F}_{{\cdot,\cdot }}\;, \label{cocyclef}
\end{equation}
as well as the normalization condition
\begin{equation}
(\epsilon \otimes 1)~{\cal F}_{{\cdot,\cdot }} = (1 \otimes
\epsilon)~{\cal F}_{{\cdot,\cdot }} = 1\;, \label{normalizationhh}
\end{equation}
with ${\cal F}_{{\cdot,\cdot }12} = {\cal F}_{{\cdot,\cdot }}\otimes
1$, ${\cal F}_{{\cdot,\cdot }23} = 1 \otimes {\cal F}_{{\cdot,\cdot
}}$ and $\Delta _{0}(a) = a \otimes 1 + 1 \otimes a$. \\
In accordance with the general  twist quantization procedure
\cite{qg1}-\cite{qg3} the algebraic sectors of all discussed below
Hopf  structures remain undeformed $(\eta_{\mu\nu} = (-,+,+,+))$
\begin{eqnarray}
&&\left[ M_{\mu \nu },M_{\rho \sigma }\right] =i\left( \eta _{\mu
\sigma }\,M_{\nu \rho }-\eta _{\nu \sigma }\,M_{\mu \rho }+\eta
_{\nu \rho }M_{\mu
\sigma }-\eta _{\mu \rho }M_{\nu \sigma }\right) \;,  \notag \\
&&\left[ M_{\mu \nu },P_{\rho }\right] =i\left( \eta _{\nu \rho
}\,P_{\mu }-\eta _{\mu \rho }\,P_{\nu }\right) \;\;\;,\;\;\; \left[
P_{\mu },P_{\nu }\right] =0\;,   \label{nnn}
\end{eqnarray}
while the   coproducts and antipodes  transform as follows
\begin{equation}
\Delta _{0}(a) \to \Delta _{\cdot,\cdot }(a) =
\mathcal{F}_{\cdot,\cdot }\circ \,\Delta _{0}(a)\,\circ
\mathcal{F}_{\cdot,\cdot }^{-1}\;\;\;,\;\;\; S_{\cdot}(a)
=u_{\cdot,\cdot }\,S_{0}(a)\,u^{-1}_{\cdot,\cdot }\;,\label{fs}
\end{equation}
where   $S_0(a) = -a$ and $u_{\cdot,\cdot }=\sum
f_{(1)}S_0(f_{(2)})$ (we use Sweedler's notation
$\mathcal{F}_{\cdot,\cdot }=\sum f_{(1)}\otimes f_{(2)}$).

Recently, in the article \cite{genpogali}, there have been
considered three (all possible) types  of  Abelian and generalized
twist factors\footnote{Indecies $k$, $l$ are fixed and different
than $i$.}, generating
the noncommutative space-time algebra of type (\ref{general})\footnote{$a\wedge b = a\otimes b - b\otimes a$.}:\\
 \begin{eqnarray}
   i)\;\;\;\mathcal{F}_{\theta_{kl},{\kappa}}  = \exp\;i\left[\frac{1}{2{\kappa}}P_k
\wedge M_{i0} + \theta_{kl}P_{k}\wedge P_{l}\right] \;, \label{rge1}
\end{eqnarray}
\begin{eqnarray}
  ii)\;\;\; \mathcal{F}_{\theta_{0i},\hat{\kappa}}  =
\exp \;i\left[\frac{1}{2\hat{\kappa}}P_0 \wedge M_{kl} +
{{\theta}_{0i}}P_{0 }\wedge P_{i}\right]  \;, \label{rge2}
\end{eqnarray}
\begin{eqnarray}
  iii)\;\;\; \mathcal{F}_{\theta_{0i},\bar{\kappa}}  =
\exp \;i\left[\frac{1}{2\bar{\kappa}}P_i \wedge M_{kl} +
{{\theta}_{0i}}P_{0 }\wedge P_{i}\right] \;. \label{rge3}
\end{eqnarray}
They lead to the following coproduct sector in the case of
deformation
 $\mathcal{U}_{\theta_{kl},\kappa}(P)$ generated by the twist
 factor (\ref{rge1})
\begin{eqnarray}
\Delta_{\theta_{kl},{\kappa}}(P_\mu)&=&\Delta
_0(P_\mu)+\sinh(\frac{1}{2 {\kappa}} P_k )\wedge
\left(\eta_{i \mu}P_0 -\eta_{0 \mu}P_i \right)\label{coa1}\\
&+&(\cosh(\frac{1}{2 {\kappa}}  P_k )-1)\perp \left(\eta_{i \mu}P_i
-\eta_{0 \mu}P_0 \right)\;,\notag
\end{eqnarray}
\begin{eqnarray}
\Delta_{\theta_{kl},{\kappa}}(M_{\mu\nu})&=&\Delta_0(M_{\mu\nu})+\frac{1}{2
{\kappa}}M_{i 0 }\wedge  \left(\eta_{\mu
k }P_\nu-\eta_{\nu k}P_\mu\right)\nonumber\\
&+&i\left[M_{\mu\nu},M_{i 0 }\right]\wedge
\sinh(\frac{1}{2 {\kappa}} P_k ) \notag \\
&-&\left[\left[%
M_{\mu\nu},M_{i 0 }\right],M_{i 0 }\right]\perp
(\cosh(\frac{1}{2 {\kappa}}  P_k  )-1) \nonumber \\
&+&\frac{1}{2 {\kappa}}M_{i 0 }\sinh(\frac{1}{2 {\kappa}} P_k )\perp
 \left(\psi_k P_i -\chi_k P_0 \right) \nonumber \\
&-&\frac{1}{2 {\kappa}} \left(\psi_k P_0 -\chi_k P_i \right)\wedge
M_{i 0 }(\cosh(\frac{1}{2 {\kappa}} P_k )-1) \notag
\end{eqnarray}
\begin{eqnarray}
&-& \theta_{k l }[(\eta _{k \mu }P_{\nu }-\eta _{k \nu }\,P_{\mu
})\otimes P_{l }+P_{k}\otimes (\eta_{l
\mu}P_{\nu}-\eta_{l \nu}P_{\mu})]\label{coa100}\\
&+& \theta_{kl}[(\eta _{l \mu }P_{\nu }-\eta _{l \nu }\,P_{\mu
})\otimes P_{k }+P_{l}\otimes (\eta_{k \mu}P_{\nu}-\eta_{k
\nu}P_{\mu})]\nonumber\\
&~~&\nonumber\\
&+& \theta_{kl}\left[\left[M_{\mu\nu},M_{i 0 }\right],P_k\right]
\perp
\sinh(\frac{1}{2 {\kappa}} P_k )P_l\notag\\
&-& \theta_{kl}\left[\left[M_{\mu\nu},M_{i 0 }\right],P_l\right]
\perp \sinh(\frac{1}{2 {\kappa}} P_k )P_k
\notag\\
&+&i\theta_{kl}\left[\left[\left[%
M_{\mu\nu},M_{i 0 }\right],M_{i 0 }\right],P_k\right] \wedge
(\cosh(\frac{1}{2 {\kappa}}  P_k  )-1)P_l
\notag\\
&-& i\theta_{kl}\left[\left[\left[%
M_{\mu\nu},M_{i 0 }\right],M_{i 0 }\right],P_l\right] \wedge
(\cosh(\frac{1}{2 {\kappa}}  P_k  )-1)P_k\;,\notag
\end{eqnarray}
with   $a\perp b=a\otimes b+b\otimes a$, $\psi_\gamma =\eta_{j
\gamma }\eta_{l i}-\eta_{i \gamma }\eta_{lj}$ and $\chi_\gamma
=\eta_{j \gamma }\eta_{k i}-\eta_{i \gamma }\eta_{k j}$. The  two
remaining Hopf structures
($\mathcal{U}_{\theta_{0i},\hat{\kappa}}(P)$ and
$\,\mathcal{U}_{\theta_{0i},\bar{\kappa}}(P)$) corresponding to the
twist factors (\ref{rge2}) and (\ref{rge3}) look  similar (see
\cite{genpogali}) to the coproducts (\ref{coa1}) and (\ref{coa100}).
For this reason they will be omitted in present  article.

Obviously, for the deformation parameter $\theta_{kl}$ approaching
zero and parameter $\kappa$ running to infinity, the above Hopf
structure becomes classical. Besides,  for fixed (different than
zero) parameter $\theta_{kl}$  and parameter $\kappa$ approaching
infinity, we get twisted (canonical) Poincare Hopf algebra provided
in \cite{3a}. Moreover, for parameter $\theta_{kl}$  running to zero
and fixed parameter $\kappa$, we recover the Lie-algebraically
deformed relativistic Hopf algebra introduced in  \cite{lie2} (see
also \cite{lie1}).

\subsection{Nonrelativistic case}

The corresponding Galilei Hopf algebras
$\,\mathcal{U}_{\xi_{kl},\lambda}(G)$,
$\,\mathcal{U}_{\xi_{0i},\hat{\lambda}}(G)$ and
$\,\mathcal{U}_{\xi_{0i},\bar{\lambda}}(G)$ can be obtained by
direct application of  twist procedure (see formula (\ref{fs})) or
by nonrelativistic contraction of deformations
$(\ref{rge1})$-$(\ref{rge3})$. They have been found in
\cite{genpogali} with the  use of contraction procedure which leads
to the following algebraic sector
\begin{eqnarray}
&&\left[\, K_{ab},K_{cd}\,\right] =i\left( \delta
_{ad}\,K_{bc}-\delta
_{bd}\,K_{ac}+\delta _{bc}K_{ad}-\delta _{ac}K_{bd}\right) \;,  \notag \\
&&\left[\, K_{ab},V_{c}\,\right] =i\left( \delta _{bc}\,V_a-\delta
_{ac}\,V_b\right)\;\; \;, \;\;\;\left[ \,K_{ab},\Pi_{c }\,\right]
=i\left( \delta _{b c }\,\Pi_{a }-\delta _{ac }\,\Pi_{b }\right) \;,
\label{nnnga}
\\
&&\left[ \,K_{ab},\Pi_{0 }\,\right] =\left[ \,V_a,V_b\,\right] =
\left[ \,V_a,\Pi_{b }\,\right] =0\;\;\;,\;\;\;\left[ \,V_a,\Pi_{0
}\,\right] =-i\Pi_a\;\;\;,\;\;\;\left[ \,\Pi_{\rho },\Pi_{\sigma
}\,\right] = 0\;,\nonumber
\end{eqnarray}
and the coalgebraic one
\begin{eqnarray}
 \Delta_{\xi_{kl},{\lambda}}(\Pi_0)&=&\Delta _0(\Pi_0) +
\frac{1}{2{{\lambda}}} \Pi_k \wedge \Pi_i\;,\label{gacoa1}\\
\Delta_{\xi_{kl},{\lambda}}(\Pi_a)&=&\Delta
_0(\Pi_a)\;\;\;,\;\;\;\Delta_{\xi_{kl},{\lambda}}(V_a)=\Delta_0(V_a)\;,\label{coa0}\\
 &~~&  \cr
\Delta_{\xi_{kl},{\lambda}}(K_{ab})&=&\Delta_0(K_{ab})+
\frac{i}{2{{\lambda}}}\left[\;K_{ab},V_i\;\right]\wedge \Pi_k +
\frac{1}{2{{\lambda}}}V_i \wedge(\delta_{ak}\Pi_b
-\delta_{bk}\Pi_a) \nonumber\\ 
&-&\xi_{k l }\left[\;(\delta_{k a}\Pi_{b }-\delta_{k b
}\,\Pi_{a})\otimes \Pi_{l }+\Pi_{k}\otimes (\delta_{l
a}\Pi_{b}-\delta_{l b}\Pi_{a})\;\right]
 \label{gacoa100}\\
 &\;&\;\;\;\;+\;\;\xi_{k l }\left[\;(\delta_{l a}\Pi_{b }-\delta_{l b }\,\Pi_{a})\otimes
\Pi_{k }+\Pi_{l}\otimes
(\delta_{ka}\Pi_{b}-\delta_{kb}\Pi_{a})\;\right] \nonumber\;,
\end{eqnarray}
where the generators $K_{ij}$, $\Pi_{\mu}$ and $V_i$ are given by
\begin{equation}
P_{0 } = \frac{{\Pi_{0 }}}{c}\;\;\;,\;\;\;P_{i } =
\Pi_{i}\;\;\;,\;\;\;M_{ij}= K_{ij}\;\;\;,\;\;\;M_{i0}= cV_i\;,
\label{contr2}
\end{equation}
and
\begin{equation}
\lambda = {\kappa}/{c}\;\;,\;\;\xi_{kl} =
\theta_{kl}\;\;(\xi_{lk}=\theta_{lk})\;\;,\;\;c - {\rm
light\;velocity} \;. \label{contr3}
\end{equation}
The two remanning cosectors for
$\mathcal{U}_{\xi_{0i},\hat{\lambda}}(G)$ and
$\,\mathcal{U}_{\xi_{0i},\bar{\lambda}}(G)$ associated with the
twist factors (\ref{rge2}) and (\ref{rge3}) look similar to the
formulas (\ref{gacoa1})-(\ref{gacoa100}) (see \cite{genpogali}) and
are omitted in the present article.

Obviously, for  deformation parameter $\xi_{kl}$  approaching zero
and parameter $\lambda$ running to infinity, the above Hopf algebra
becomes classical. Besides,  for fixed parameter $\xi_{kl}$ and
parameter $\lambda$ approaching infinity, we get twisted (canonical)
Galilei Hopf algebra provided in \cite{dasz}. Moreover, for
parameter $\xi_{kl}$  running to zero and fixed parameter $\lambda$,
we recover the Lie-algebraically deformed nonrelativistic Hopf
algebra introduced in \cite{dasz} as well.

\section{Generalized twist deformations of Poincare and
Ga-
lilei (dual) quantum groups}

\subsection{Relativistic case}

Let us now turn to the Poincare quantum groups ${\mathcal
P}_{\theta_{kl},\kappa}$, ${\mathcal P}_{\theta_{0i},\hat{\kappa}}$
and ${\mathcal P}_{\theta_{0i},\bar{\kappa}}$ dual to the
relativistic Hopf algebras  $i)$, $ii)$ and $iii)$ provided in
pervious section.

It is well known that such  structures can be obtained with use of
so-called FRT procedure \cite{frt}. Hence, in a first step of our
algorithm we introduce the quantum $R$-matrices associated with
considered Poincare  groups. They satisfy so-called quantum
Yang-Baxter equation (QYBE)
\begin{equation}
R_{12}R_{13}R_{23} = R_{23}R_{13}R_{12}\;\;\;;\;\;\;R_{12} =
R_\alpha \otimes R_\beta \otimes 1\;\;{\rm if}\;R= R_\alpha \otimes
R_\beta\;\;{\rm etc}\;, \label{qybe}
\end{equation}
and in the case of twisted algebras, they take the form (see
\cite{qg1}-\cite{qg3})
\begin{equation}
R_{\cdot,\cdot}
=\mathcal{F}_{\cdot,\cdot}^{\,T}\cdot\mathcal{F}_{\cdot,\cdot
}^{-1}=\exp \left(-{2}i\,r_{\cdot,\cdot}\right)\;\;\;,\;\;\;
(a\otimes b)^{T}=b\otimes a\;.  \label{quantumR}
\end{equation}
Particulary, in the standard matrix representation of the Poincare
generators
\begin{equation}
(M_{\mu \nu })_{\ B}^{A}=\delta _{\ \mu }^{A}\eta _{\nu B }-\delta
_{\ \nu }^{A}\eta _{\mu B }\;\;\;,\;\;\; (P_{\mu })_{\ B}^{A}=\delta
_{\ \mu }^{A}\delta _{\ B}^{4}\;, \label{rep}
\end{equation}
the $R$-matrices associated with twist factors
(\ref{rge1})-(\ref{rge3}) look as follows\footnote{For  matrix
representation (\ref{rep}) we have $P_\mu P_\nu = P_\mu M_{\rho\tau}
= 0$ in the case of
commuting generators $P_\mu$ and $M_{\rho\tau}$.}:
 \begin{eqnarray}
   i)\;\;\;R_{\theta_{kl},{\kappa}}  = 1\otimes 1 + i\left[\;\frac{1}{{\kappa}}P_k
\wedge M_{i0} + 2\theta_{kl}P_{k}\wedge P_{l}\;\right] \;,
\label{quantumr1}
\end{eqnarray}
\begin{eqnarray}
  ii)\;\;\; R_{\theta_{0i},\hat{\kappa}}  =
1\otimes 1+i\left[\;\frac{1}{\hat{\kappa}}P_0 \wedge M_{kl} +
2{{\theta}_{0i}}P_{0 }\wedge P_{i}\;\right]  \;, \label{quantumr2}
\end{eqnarray}
and
\begin{eqnarray}
  iii)\;\;\; R_{\theta_{0i},\bar{\kappa}}  =1\otimes 1+
i\left[\;\frac{1}{\bar{\kappa}}P_i \wedge M_{kl} +
2{{\theta}_{0i}}P_{0 }\wedge P_{i}\;\right] \;, \label{quantumr3}
\end{eqnarray}
respectively.

In the second step of FRT procedure we introduce the following
$5\times 5$ - matrices
\begin{equation}
{{T}}_{\ B}^{A}=\left[
\begin{array}{cc}
{\Lambda} _{\ \nu }^{\mu } & {a}^{\mu } \\
0 & 1
\end{array}
\right] \;,  \label{tmatrix}
\end{equation}
where ${\Lambda} _{\ \nu }^{\mu }$ parametrizes the quantum Lorentz
rotation and ${a}^{\mu }$ denotes quantum translations, such that
\begin{eqnarray}
< \Lambda^{\mu}_{~\nu},M^{\alpha\beta} > \,= \left(\eta^{\alpha\mu}
\delta^{\beta}_{~\nu} - \eta^{\beta\mu} \delta^{\alpha}_{~\nu}
\right)\;\;\;,\;\;\; < a^{\mu},P_{\nu} > \,=
\delta^{\mu}_{~\nu}\;.\label{dualnosc}
\end{eqnarray}
Consequently, the algebraic part of dual Poincare group is described
by  so-called RTT relation
\begin{equation}
{R}_{\cdot,\cdot}{{T}}_{1}{{T}}_{2}={{T}}_{2}{{T}}_{1}{R}_{\cdot,\cdot}\;,
\label{rtt}
\end{equation}
while the composition law for the coproduct remains classical
\begin{equation}
\Delta ({{T}}_{\ B}^{A})={{T}}_{\ C}^{A}\otimes {{T}}_{\ B}^{C}\;,
\label{prymityw}
\end{equation}
with ${{T}}_{1}={{T}}\otimes 1$, ${{T}}_{2}=1\otimes {{T}}$ and
quantum ${R}$-matrix given in the representation
(\ref{rep}).\newline Using results presented in \cite{3c} and
\cite{lie2} after insertion of (\ref{quantumr1}), one can write the
relations (\ref{rtt}) in terms of the operator basis $({\Lambda} _{\
\nu }^{\mu },{a}^{\mu })$,
 as follows\footnote{Due the linear form of matrices
(\ref{quantumr1})-(\ref{quantumr3}), one can write the commutation
relations for dual groups $\mathcal{P}_{\cdot,\cdot}$ as the sum of
commutators provided in  articles \cite{3c} and \cite{lie2}.}
\begin{eqnarray}
&i)\;\;\;&\left[\; {a}^{\mu },{a}^{\nu }\;\right]
=\frac{i}{\kappa}\delta^{\nu k } ( \delta _{\ i }^{\mu }{a}_{0
}-\delta _{\ 0 }^{\mu }{a}_{i }) + \frac{i}{\kappa}\delta^{\mu k}(
\delta _{\ 0}^{\nu }{a}_{i }-\delta _{\ i }^{\nu
}{a}_{0 }) +  \nonumber\\
&&~~~~~~~~~~~-\;i\theta_{kl }( {\Lambda} _{\ k }^{\mu }\,{\Lambda}
_{\ l }^{\nu }-\delta _{\ k }^{\mu }\,\delta _{\ l }^{\nu })
-i\theta_{lk }( {\Lambda} _{\ l }^{\mu }\,{\Lambda} _{\ k }^{\nu
}-\delta _{\ l }^{\mu }\,\delta _{\ k }^{\nu })
\;,\nonumber\\
&&\left[\; {a}^{\mu },{\Lambda} _{\ \rho }^{\nu }\;\right]
=\frac{i}{\kappa}{\Lambda} _{\ k }^{\mu }( \eta _{0 \rho }{\Lambda}
_{\ i }^{\nu }-\eta _{i \rho }{\Lambda} _{\ 0 }^{\nu })
+\frac{i}{\kappa}\delta ^{\mu k }( \delta _{\ 0 }^{\nu }{\Lambda}
_{i \rho }-\delta _{\ i
}^{\nu }{\Lambda} _{0 \rho }) \;,\label{rel1a} \\
&\;&\cr &&\left[\; {\Lambda} _{\ \nu }^{\mu },{\Lambda} _{\ \tau
}^{\rho }\;\right] =0\;. \nonumber
\end{eqnarray}
Similarly, if we use second $R$-matrix
 (\ref{quantumr2}) we get
\begin{eqnarray}
&ii)\;\;&\left[\; {a}^{\mu },{a}^{\nu }\;\right] =\frac{i}{\hat
\kappa}\delta _{\ 0 }^{\nu } ( \delta _{\ k }^{\mu }{a}_{l }-\delta
_{\ l }^{\mu }{a}_{k }) + \frac{i}{\hat \kappa}\delta _{\ 0 }^{\mu
}( \delta _{\ l }^{\nu }{a}_{k }-\delta _{\ k }^{\nu
}{a}_{l }) +   \nonumber \\
&&~~~~~~~~~~~+\;i\theta_{0i }( {\Lambda} _{\ 0 }^{\mu }\,{\Lambda}
_{\ i }^{\nu }-\delta _{\ 0 }^{\mu }\,\delta _{\ i }^{\nu })
+i\theta_{i0 }( {\Lambda} _{\ i }^{\mu }\,{\Lambda} _{\ 0 }^{\nu
}-\delta _{\ i }^{\mu }\,\delta _{\ 0 }^{\nu })
\;,\nonumber\\
&&\left[\; {a}^{\mu },{\Lambda} _{\ \rho }^{\nu }\;\right]
=\frac{i}{\hat\kappa}{\Lambda} _{\ 0 }^{\mu }( \eta _{l \rho
}{\Lambda} _{\ k }^{\nu }-\eta _{k \rho }{\Lambda} _{\ l }^{\nu })
+\frac{i}{\hat \kappa}\delta _{\ 0 }^{\mu }( \delta _{\ l }^{\nu
}{\Lambda} _{k \rho }-\delta _{\ k }^{\nu }{\Lambda} _{l \rho }) \;,
\label{rel2b} \\
&\;&\cr &&\left[\; {\Lambda} _{\ \nu }^{\mu },{\Lambda} _{\ \tau
}^{\rho }\;\right] =0\;. \nonumber
\end{eqnarray}
In the third  case  (\ref{quantumr3}) one obtains
\begin{eqnarray}
&iii)\;&\left[\; {a}^{\mu },{a}^{\nu }\;\right]
=\frac{i}{\bar\kappa}\delta ^{\nu i } ( \delta _{\ k }^{\mu }{a}_{l
}-\delta _{\ l }^{\mu }{a}_{k }) + \frac{i}{\bar \kappa}\delta^{\mu
i}( \delta _{\ l }^{\nu }{a}_{k }-\delta _{\ k }^{\nu
}{a}_{l })  +  \nonumber \\
&&~~~~~~~~~~~+\;i\theta _{0i }( {\Lambda} _{\ 0 }^{\mu }\,{\Lambda}
_{\ i }^{\nu }-\delta _{\ 0 }^{\mu }\,\delta _{\ i }^{\nu
})+i\theta_{i0 }( {\Lambda} _{\ i }^{\mu }\,{\Lambda} _{\ 0 }^{\nu
}-\delta _{\ i }^{\mu }\,\delta _{\ 0 }^{\nu })\;,\nonumber\\
&&\left[\; {a}^{\mu },{\Lambda} _{\ \rho }^{\nu }\;\right]
=\frac{i}{\bar \kappa}{\Lambda} _{\ i }^{\mu }( \eta _{l\rho
}{\Lambda} _{\ k }^{\nu }-\eta _{k \rho }{\Lambda} _{\ l }^{\nu })
+\frac{i}{\bar \kappa}\delta ^{\mu i}( \delta _{\ l }^{\nu
}{\Lambda} _{k \rho }-\delta _{\ k}^{\nu }{\Lambda} _{l \rho }) \;,
\label{rel3c} \\
&\;&\cr &&\left[\; {\Lambda} _{\ \nu }^{\mu },{\Lambda} _{\ \tau }^{
\rho }\;\right] =0\;.\nonumber
\end{eqnarray}
 Besides, inserting (\ref{tmatrix}) in the formula
(\ref{prymityw}) we get the well-known form of the  coproducts
\begin{equation}
\Delta_{\cdot,\cdot} \,({\Lambda} _{\ \nu }^{\mu })={\Lambda} _{\
\rho }^{\mu }\otimes {\Lambda} _{\ \nu }^{\rho }\;\;\;,\;\;\;
\Delta_{\cdot,\cdot} ({a}^{\mu })={\Lambda} _{\ \nu }^{\mu }\otimes
{a}^{\nu }+{a}^{\mu }\otimes 1\;. \label{dualcopro}
\end{equation}
It should be also noted that all above relations can be supplemented
by the classical antipode
\begin{equation}
S_{\cdot,\cdot}({\Lambda} _{\ \nu }^{\mu })={\Lambda} _{\ \nu }^{\mu
}\;\;\;,\;\;\; S_{\cdot,\cdot}({a}^{\mu })=-{\Lambda} _{\ \nu }^{\mu
}\,{a}^{\nu }\;, \label{antipode}
\end{equation}
and the counit
\begin{equation}
\epsilon_{\cdot,\cdot}({\Lambda}^\mu_{\ \nu})=\delta^\mu_{\
\nu}\;\;\;,\;\;\;\epsilon_{\cdot,\cdot}({a}^\mu)=0\;.
\end{equation}
In such a way we get three types of (dual) quantum groups ${\mathcal
P}_{\theta_{kl},\kappa}$, ${\mathcal P}_{\theta_{0i},\hat{\kappa}}$
and ${\mathcal P}_{\theta_{0i},\bar{\kappa}}$,
 equipped with the following *-involution
\begin{equation}
({a}^{\mu })^{\ast }={a}^{\mu }\;\;\;,\;\;\; ({\Lambda} _{\ \nu
}^{\mu })^{\ast }={\Lambda} _{\ \nu }^{\mu }\;. \label{involution}
\end{equation}

Obviously, for parameters $\theta_{kl}$ and $\theta_{0i}$
approaching zero, and parameters $\kappa$, $\hat\kappa$ and
${\bar\kappa}$ running to infinity, the  above deformations
disappear. Besides,  for fixed (different than zero) parameters
$\theta_{kl}$ and $\theta_{0i}$, and parameters $\kappa$,
$\hat\kappa$ and $\bar\kappa$ approaching infinity, we get twisted
Poincare group provided in \cite{3c}. Moreover, for parameters
$\theta_{kl}$ and $\theta_{0i}$ running to zero, and fixed
parameters $\kappa$, $\hat\kappa$ and $\bar\kappa$, we recover the
Lie-algebraically deformed dual Hopf structures introduced in
\cite{lie1}.

\subsection{Nonrelativistic case}

The nonrelativistic counterparts of dual quantum groups presented in
pervious subsection  can be get by:\\

a) application of already mentioned FRT procedure,\\
\\
or\\

b) nonrelativistic contractions of  Hopf structures
(\ref{rel1a})-(\ref{rel3c}).\\
\\
Here, we choose the option b). Consequently,  we perform the
contraction limit of dual quantum groups ${\mathcal
P}_{\theta_{kl},\kappa}$, ${\mathcal P}_{\theta_{0i},\hat{\kappa}}$
and ${\mathcal P}_{\theta_{0i},\bar{\kappa}}$ in two steps. Firstly,
we rewrite the Poincare generators ${\Lambda} _{\ \nu }^{\mu }$ and
${a}^{\mu }$ in terms of Galileian rotations ${R} _{\ j }^{i }$,
boosts $v^i$ and translations $(\tau,b^i)$ \cite{gg}
\begin{eqnarray}
&& {\Lambda} _{\ 0 }^{0 } = \left(
1+\frac{\overline{v}^2}{c^2}\right)^{\frac{1}{2}}\;,
\label{zadrugizm1}\\
&~~&  \cr && {\Lambda} _{\ 0 }^{i } = \frac{v^i}{c}\;,\\
&~~&  \cr
&& {\Lambda} _{\ i }^{0 }= \frac{v^k{R} _{\ i}^{k }}{c}\;,\\
&~~&  \cr && {\Lambda} _{\ i }^{k } = \left(\delta _{\ l }^{k
}+\left(\left(
1+\frac{\overline{v}^2}{c^2}\right)^{\frac{1}{2}}-1\right)
\frac{v^kv^l}{\overline{v}^2}\right){R} _{\ i }^{l } \;,\\
&~~&  \cr &&a^i = b^i\;\;\;,\;\;\;a^0 = c\tau\;. \label{zadrugizm7}
\end{eqnarray}
 Besides, we  rescale the deformation parameters
$\theta_{kl}, \theta_{0i}, \kappa, \hat{\kappa}$ and $\bar\kappa$ as
follows (see \cite{genpogali})
\begin{eqnarray}
\xi_{kl} = \theta_{kl}\;\;,\;\;\xi_{0i} = \theta_{0i}/c
\;\;,\;\;\lambda = {\kappa}/{c}\;\;,\;\;{\hat \lambda} = {\hat
\kappa}{c} \;\;,\;\; {\bar \lambda} = {\bar \kappa} \;.
\label{zadrugizm750}
\end{eqnarray}
Then, we rewrite the commutation relations
(\ref{rel1a})-(\ref{rel3c}) in terms of (new) generators
(\ref{zadrugizm1})-(\ref{zadrugizm7}) and deformation parameters
(\ref{zadrugizm750}). Finally,  we take the contraction limit $c \to
\infty$ and, in such a way, we get three Galilei quantum groups
${\mathcal G}_{\xi_{kl},\lambda}$, ${\mathcal
G}_{\xi_{0i},\hat{\lambda}}$ and ${\mathcal
G}_{\xi_{0i},\bar{\lambda}}$ dual to the Galilei Hopf algebras
$\,\mathcal{U}_{\xi_{kl},\lambda}(G)$,
$\,\mathcal{U}_{\xi_{0i},\hat{\lambda}}(G)$ and
$\,\mathcal{U}_{\xi_{0i},\bar{\lambda}}(G)$ provided in
\cite{genpogali}. They take the form:
\begin{eqnarray}
&i)\;\;\;&[\; {b}^{m },{b}^{n }\;] =\frac{i}{{\lambda}} \tau( \delta
_{\ k}^{n}\delta _{\ i}^{m }\,-\delta _{\ i }^{n }\delta _{\ k}^{m
})- i\xi_{kl }( {R} _{\ k}^{m }\,{R} _{\ l }^{n }-\delta _{\ k }^{m
}\,\delta _{\ l }^{n })+\nonumber\\
&& \;\;\;\;\;\;\;\;\;\;\;\;\;\;\;\;-i\;\xi_{lk }( {R} _{\ l}^{m
}\,{R} _{\ k }^{n }-\delta _{\ l }^{m }\,\delta _{\ k }^{n })
\;,\nonumber\\
&~~&  \cr &&[\; {v}^{n },{b}^{m }\;]=\frac{i}{{\lambda}}( {R} _{\ k
}^{m }{R} _{\ i }^{n }+\delta _{\ k }^{m }\delta _{\ i }^{n }) \;,
\label{perun}\\
&~~&  \cr &&[\; b^m,{R} _{\ q }^{p}\;]= [\; {{\tau}},b^{m }\;]=[\;
{{\tau}},v^{m }\;] =  [\; v^m,v^{n }\;]=0\;,\nonumber
\\&~~& \cr &&[\; {R} _{\ n }^{m },{R} _{\ q }^{p}\;]=[\; {v}^m,{R}
_{\ q }^{p }\;] =[\; {{\tau}},{R} _{\ n }^{m }\;]  =0\;, \nonumber
\end{eqnarray}
\begin{eqnarray}
&ii)\;\;\;&[\; {{\tau}},b^{m }\;] =\frac{i}{{\hat\lambda}} ( \delta
_{\ k }^{m }\, b_{l }-\delta _{\ l}^{m
}\,{b} _{k }) + i{\hat \xi}_{0i}(R^{m}_{\ i} + \delta^{m}_{\ i}) \;, \nonumber\\
&~~&  \cr &&[\; {{\tau}},v^{m }\;] = \frac{i}{{\hat\lambda}} (
\delta _{\ k }^{m }\, v_{l }-\delta
_{\ l}^{m }\,{v} _{k }) \;,\nonumber\\
&~~&  \cr &&[\; {{\tau}},{R} _{\ n }^{m}\;]
=\frac{i}{{\hat\lambda}}( \delta _{ln}{R} _{\ k }^{m }-\delta _{k n
}{R} _{\ l }^{m }) + \frac{i}{{\hat\lambda}}( \delta _{\ k }^{m }{R}
_{ l n }-\delta _{\
l }^{m }{R} _{ k n })\;,\nonumber\\
&~~&  \cr &&[\; b^m,{R} _{\ q}^{p}\;]=\frac{i}{{\hat\lambda}}v^{m }(
\delta _{lq}{R} _{\ k }^{p }-\delta _{k q }{R} _{\ l }^{p })
\;,\label{nadnarodslawski}\\
&~~&  \cr &&[\; {b}^{m },{b}^{n }\;]=i{{ \xi}_{0i }}( v^{m }\,{R}
_{\ i }^{n }-{R} _{\ i }^{m}v^n )\;,\nonumber
\\
&~~&  \cr &&[\; {b}^{m },{v}^{n }\;] = [\; v^m,v^{n }\;]=[\; {R} _{\
n }^{m },{R} _{\ q }^{p}\;]=[\; {v}^m,{R} _{\ q }^{p }\;] =0\;,
\nonumber
\end{eqnarray}
and
\begin{eqnarray}
&iii)\;\;\;& [\; {b}^{m },{b}^{n }\;]= \frac{i}{{\bar\lambda}}
\delta^{ni}( \delta _{\ k}^{m }b_l\,-\delta _{\ l }^{m }b_k)+
\frac{i}{{\bar\lambda}} \delta^{mi}( \delta _{\ l}^{n }b_k\,-\delta
_{\ k }^{n }b_l)+
\nonumber\\
&& \;\;\;\;\;\;\;\;\;\;\;\;\;\;\;\;+i\;{\xi}_{0i } ( v^{m }\,{R} _{\
i }^{n }-{R} _{\ i }^{m}v^n )\;,
\nonumber\\
&~~&  \cr&&[\; {{\tau}},b^{m }\;]=i{ \xi}_{0i}(R^{m}_{\ i} + \delta^{m}_{\ i}) \;, \nonumber\\
&~~&  \cr  &&[\; {b}^{m },{v}^{n
}\;]=\frac{i}{{\bar\lambda}}\delta^{mi}( \delta _{\ l }^{n }v^k-
\delta _{\ k }^{n }v^l)
\;,\label{sssystem}\\
&~~& \cr &&[\; b^m,{R} _{\ q }^{p}\;]= \frac{i}{{\bar\lambda}}{R}
_{\ i}^{m} (\delta_{lq}{R} _{\ k}^{p} - \delta_{kq}{R} _{\ l}^{p}) +
\frac{i}{{\bar\lambda}}{\delta}^{mi}(\delta _{\ l}^{p}{R} _{kq} -
\delta _{\ k}^{p}{R} _{lq})
\;,\nonumber\\
&~~&  \cr && [\; {{\tau}},v^{m }\;]=[\; {{\tau}},{R} _{\ n}^{m}\;]=
[\; v^m,v^{n }\;]=[\; {R} _{\ n }^{m},{R} _{\ q }^{p}\;]=[\;
{v}^m,{R} _{\ q }^{p}\;] =0\;, \nonumber
\end{eqnarray}
respectively. The corpoducts, counits and antipodes remain
classical.

Obviously, for parameters $\xi_{kl}$ and $\xi_{0i}$ approaching
zero, and parameters $\lambda$, $\hat\lambda$ and ${\bar\lambda}$
running to infinity, the  above deformations disappear. Besides, for
fixed parameters $\xi_{kl}$ and $\xi_{0i}$, and parameters
$\lambda$, $\hat\lambda$ and $\bar\lambda$ approaching infinity, we
get twisted nonrelativistic quantum groups introduced in
\cite{dualdasz}. Moreover, for parameters $\xi_{kl}$ and $\xi_{0i}$
running to zero, and fixed parameters $\lambda$, $\hat\lambda$ and
$\bar\lambda$, we recover the Lie-algebraically deformed Galilei
Hopf structures introduced as well in \cite{dualdasz}.

\section{Final remarks}

In this article we provide six quantum groups ${\mathcal
P}_{\theta_{kl},\kappa}$, ${\mathcal P}_{\theta_{0i},\hat{\kappa}}$,
 ${\mathcal P}_{\theta_{0i},\bar{\kappa}}$ and
${\mathcal G}_{\xi_{kl},\lambda}$, ${\mathcal
G}_{\xi_{0i},\hat{\lambda}}$, ${\mathcal
G}_{\xi_{0i},\bar{\lambda}}$ dual to the (generalized) Poincare Hopf
algebras $\,\mathcal{U}_{\theta_{kl},\kappa}(P)$,
$\,\mathcal{U}_{\theta_{0i},\hat{\kappa}}(P)$,
$\,\mathcal{U}_{\theta_{0i},\bar{\kappa}}(P)$ and twist deformed
Galilei Hopf structures $\,\mathcal{U}_{\xi_{kl},\lambda}(G)$,
$\,\mathcal{U}_{\xi_{0i},\hat{\lambda}}(G)$,
$\,\mathcal{U}_{\xi_{0i},\bar{\lambda}}(G)$, respectively. The
relativistic quantum groups ${\mathcal P}_{\theta_{kl},\kappa}$,
${\mathcal P}_{\theta_{0i},\hat{\kappa}}$ and ${\mathcal
P}_{\theta_{0i},\bar{\kappa}}$ were obtained with use of FRT
procedure \cite{frt}, while their nonrelativistic counterparts
${\mathcal G}_{\xi_{kl},\lambda}$, ${\mathcal
G}_{\xi_{0i},\hat{\lambda}}$ and ${\mathcal
G}_{\xi_{0i},\bar{\lambda}}$ have been provided by the application
of
 well-known nonrelativistic contraction scheme
\cite{cont1}-\cite{cont3}.

It should be noted that obtained  results can be extended in various
ways. First of all, one can find with the use of Heisenberg Double
procedure \cite{qg1}-\cite{qg3}, the relativistic and
nonrelativistic phase spaces corresponding to the above Hopf
structures. Besides, it seems quite interesting to ask about basic
physical models associated with  presented here Hopf algebras and
their dual quantum groups. Such investigations have been already
initiated in the context of classical and quantum  mechanics  as
well as field theory models  (for twist deformations
(\ref{noncomm})-(\ref{noncomm2}) see \cite{przeglad} and references
therein). The works in these directions already started and are in
progress.

\section*{Acknowledgments}
The author would like to thank J. Lukierski and M. Woronowicz
for valuable comments.\\
This paper has been financially supported by Polish Ministry of
Science and Higher Education grant NN202318534.

\eject

\thispagestyle{empty}

\begin{picture}(0,0)

\put(15,-27){\framebox(120,75)} \put(52,30){Classical}
\put(35,15){Poincare} \put(55,0){Hopf algebra}
\put(45,-15){$\mathcal{U}_{0}(P)$}

\put(15,-210){\framebox(120,75)} \put(52,-150){Twisted}
\put(35,-165){Poincare}\put(55,-180){Hopf algebra}
\put(45,-195){$\mathcal{U}_{\cdot,\cdot}(P)$}

\put(15,-404){\framebox(120,75)} \put(52,-344){Twisted}
\put(35,-359){Galilei} \put(55,-374){Hopf algebra}
\put(45,-389){$\mathcal{U}_{\cdot,\cdot}(G)$}

\put(15,-615){\framebox(120,75)} \put(52,-558){Classical}
\put(35,-574){Galilei} \put(55,-588){Hopf
algebra}\put(45,-605){$\mathcal{U}_{0}(G)$}

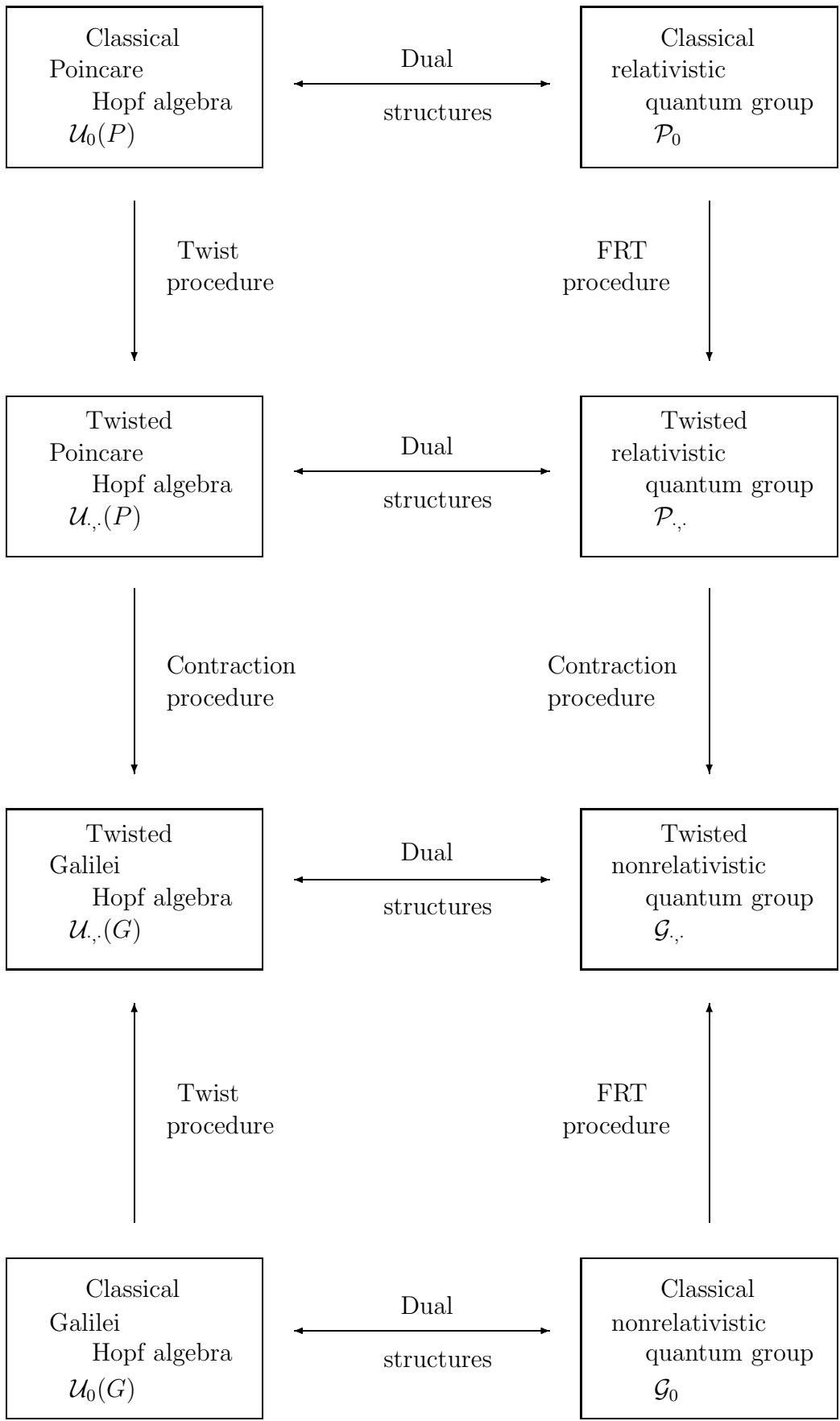
\put(30,-650){Figure 1: Twisting, contraction and duality
procedures.}

\put(285,-27){\framebox(120,75)} \put(322,30){Classical}
\put(299,15){relativistic} \put(315,0){quantum group}
\put(319,-15){$\mathcal{P}_0$}

\put(285,-210){\framebox(120,75)} \put(322,-150){Twisted}
\put(299,-165){relativistic}\put(315,-180){quantum group}
\put(319,-195){$\mathcal{P}_{\cdot,\cdot}$}

\put(285,-404){\framebox(120,75)} \put(322,-344){Twisted}
\put(299,-359){nonrelativistic} \put(315,-374){quantum group}
\put(319,-389){$\mathcal{G}_{\cdot,\cdot}$}

\put(285,-615){\framebox(120,75)} \put(322,-558){Classical}
\put(299,-574){nonrelativistic} \put(315,-588){quantum
group}\put(319,-605){$\mathcal{G}_{0}$}

\multiput(150,12)(0,0){3}{\vector(1,0){120}}
\multiput(150,-170)(0,0){3}{\vector(1,0){120}}
\multiput(150,-362)(0,0){3}{\vector(1,0){120}}
\multiput(150,-574)(0,0){3}{\vector(1,0){120}}

\multiput(150,12)(0,0){3}{\vector(-1,0){0}}
\multiput(150,-170)(0,0){3}{\vector(-1,0){0}}
\multiput(150,-362)(0,0){3}{\vector(-1,0){0}}

\multiput(150,-574)(0,0){3}{\vector(-1,0){0}}

\multiput(75,-43)(0,0){3}{\vector(0,-1){75}}
\multiput(75,-225)(0,0){3}{\vector(0,-1){87}}
\multiput(75,-523)(0,0){3}{\vector(0,1){103}}

\multiput(345,-225)(0,0){3}{\vector(0,-1){87}}
\multiput(345,-523)(0,0){3}{\vector(0,1){103}}
\multiput(345,-43)(0,0){3}{\vector(0,-1){75}}

\put(200,20){$\rm Dual$}\put(192,-5){$\rm structures$}
\put(200,-162){$\rm Dual$}\put(192,-187){$\rm structures$}
\put(200,-353){$\rm Dual$}\put(192,-378){$\rm structures$}
\put(200,-566){$\rm Dual$}\put(192,-591){$\rm structures$}

\put(95,-70){$\rm Twist$}\put(90,-85){$\rm procedure$}
\put(95,-466){$\rm Twist$}\put(90,-481){$\rm procedure$}

\put(90,-265){$\rm Contraction$}\put(90,-280){$\rm procedure$}
\put(269,-265){$\rm Contraction$}\put(269,-280){$\rm procedure$}

\put(292,-466){$\rm FRT$}\put(276,-481){$\rm procedure$}
\put(292,-70){$\rm FRT$}\put(276,-85){$\rm procedure$}

\end{picture}

\end{document}